\documentclass[floatfix,superscriptaddress,showpacs,twocolumn,amssymb,amsfonts,amsmath,prl,aps]{revtex4}
\usepackage{graphicx,epsfig}
\usepackage{upgreek}
\usepackage{multirow}
\usepackage{textcomp}     
\usepackage{color}

\usepackage{tabularx}
\newcolumntype{C}[1]{>{\centering\arraybackslash}p{#1}}
\newcommand{\be}{\begin{equation}}
\newcommand{\ee}{\end{equation}}
\newcommand{\av}[1]{\langle #1 \rangle}

\begin{document}
\title{Scaling theory for spontaneous imbibition in random networks of elongated pores
}

\author{Zeinab Sadjadi}
\affiliation{Theoretical Physics, Saarland University, 
66041 Saarbr\"ucken (Germany)}

\author{Heiko Rieger}
\affiliation{Theoretical Physics, Saarland University, 
66041 Saarbr\"ucken (Germany)}

\date{\today}

\begin{abstract}
We present a scaling theory for the long time behavior of spontaneous
imbibition in porous media consisting of interconnected pores with a
large length-to-width ratio. At pore junctions the meniscus
propagation in one or more branches can come to a halt when the
Laplace pressure of the meniscus exceeds the hydrostatic pressure
within the junction. We derive the scaling relations for the emerging
arrest time distribution and show that the average front width is
proportional to the height, yielding a roughness exponent of exactly
$\beta=1/2$ and explaining recent experimental results for nano-porous
Vycor glas (NVG). Extensive simulations of a pore network model
confirm these predictions.
\end{abstract}


\maketitle
 
The dynamics of imbibition front of an invading fluid in disordered
media has attracted substantial scientific attention, from statistical
physics \cite{Kardar98,Krug97,Alava04} to material science
\cite{Courbin07}. Besides its scientific interest, understanding the
mechanisms of imbibition in a porous matrix is of importance
in industrial processes such as oil recovery, food processing,
impregnation, chromatography, and agriculture
\cite{Alava04,Sahimi1993,Halpinhealy1995,Hinrichsen2000}.

During imbibition the liquid-gas interface advances and broadens. The
time evolution of the invading front follows simple scaling laws,
which are independent of the micro-structure and the details of the
fluid
\cite{Herminghaus02,Buldyrev92,Horvath95,Hernandez01,Planet07,Dube07}, 
reminiscent of the universality of critical phenomena.  Various
physical aspects are involved in the imbibition of a liquid inside a
porous matrix, such as viscous drag, capillarity, gravity and volume
conservation. The often complex topology of the porous matrix induces
local fluctuations in capillary pressures at the interface as well as
hydraulic permeabilities in the bulk. Despite these complexities, the
average position of the front $ \langle h(t)\rangle$ during a purely spontaneous
imbibition evolves as $\langle h(t)\rangle {\sim} t^{1/2}$, known as Lucas-Washburn
law \cite{Alava04,Lucas18, Washburn21}. This scaling behavior is valid
down to nanoscopic pore scales
\cite{Dimitrov07,Gruener09,Reyssat08}.

While the invading front exhibits a common slow-broadening dynamics
for a wide range of materials \cite{Hernandez01,
Herminghaus02,Buldyrev92, Horvath95}, the results of recent
experiments on nano-porous Vycor glass (NVG) reveal that the
roughening dynamics might depend on the micro-structure
\cite{Gruener12}.  The elongation of pores, quantitatively described
by their length-to-width radius, appears to play an important role and
two extreme limits can be distinguished: (i) \emph{short pores with
comparable length and diameter}. In materials like paper, sand,
randomly packed glass beads, etc., where the pore space is highly
interconnected \cite{Hernandez01, Herminghaus02,Buldyrev92,
Horvath95}, neighboring menisci coalesce, a continuous imbibition
front forms and an effective surface tension emerges. Due to the
latter menisci advancement is spatially highly correlated
\cite{Dube00a}, which reduces the height fluctuations of the front by
limiting menisci advancement beyond the average front position and
drawing forward the menisci lagging behind. This forms a continuous
liquid-gas interface and smoothens the front. (ii) \emph{elongated
pores}. Other porous materials like rock, soil, and porous glasses
consist of sponge-like topologies with reduced connectivity and
elongated pores \cite{Sahimi1993,Song00,Gelb98}. For example, NVG is a
silica substrate with an interconnected network of long cylindrical
pores with characteristic radii of $3{-}5$ nm. In
Ref.~\cite{Gruener12}, an anomalously fast interface roughening has
been observed, representing a new universality class
for spontaneous imbibition, emerging for large pore aspect
ratio. Here, the interface is not able to establish an effective
surface tension, leading to strong height fluctuations of the menisci.

\begin{figure}[b] \center
\includegraphics[width=1\linewidth]{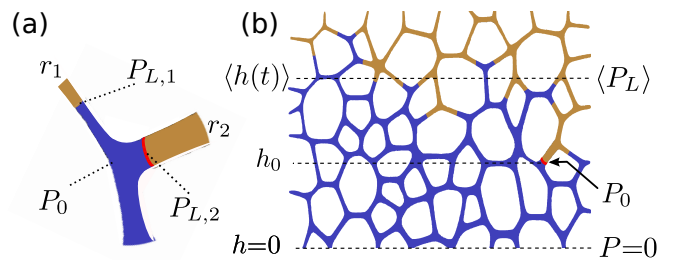}
\caption{
Sketch of a junction (a) in a pore network with elongated pores (b).
$r_i$ and $P_{L,i}$ denote the radius and Laplace pressure,
respectively, in pore $i$, and $p_0$ denotes the hydrostatic pressure
in the junction. In (b) $\langle h(t)\rangle$ and $\langle P_L\rangle$
denote the average height at time $t$ and the average Laplace
pressure.
} 
\label{fig0}
\end{figure}

In this letter we present a scaling theory for spontaneous imbibition
in porous media consisting of a network of interconnected elongated
pores (Fig~.\ref{fig0}). It is based on the observation that at pore
junctions the meniscus propagation in the branch with the larger
radius can come to a halt when the Laplace pressure of the meniscus
exceeds the hydrostatic pressure within the junction. This leads to
the emergence of voids behind the invasion front and concomitantly to
anomalously fast front broadening as observed experimentally in NVG
\cite{Gruener12}. It is predicted that the distribution of the
meniscus arrest times scales with the square of the height of the
meniscus, which implies that the ratio of the average invasion front
width and the average front height is independent of time. This
implies that roughening is maximal with an exponent $\beta=1/2$,
establishing a universality class different from those known
before for spontaneous imbibition \cite{Alava04}. We then test these
predictions in extensive simulations of a pore network model.

We analyze spontaneous imbibition of a wetting liquid in a porous
medium similar to porous glasses, which consists of a network of
elongated pores with a length-to-width ratio of the order of 10,
i.e. elongated, cylinder-like pores with random radii interconnected
at pore junctions as sketched in Fig. 1. The bottom pores are
connected to a liquid reservoir with pressure $p=0$. We assume that in
each pore a liquid-gas interface forms, denoted as meniscus, that
gives rise to a Laplace-pressure $P_L{=}{-}2\sigma/r$, where $\sigma$
is the surface tension of the liquid and $r$ the pore radius. If the
pore radii vary between $r_{\rm min}$ and $r_{\rm max}$, the average
radius is denoted by $\langle r\rangle$.  Then, on large scales, the
average height is expected to vary as $d/dt\,\langle
h(t)\rangle=-\langle P_L\rangle/\langle h(t)\rangle$, which implies
the Lucas-Washburn law $\langle h(t)\rangle\propto t^{1/2}$.

Consider now a junction at height $h_0$, where a pore branches into
two (see Fig.1a). One branch has radius $r_1$, the other $r_2>r_1$,
yielding the Laplace pressures $P_{L,i}{=-}2\sigma/r_i$. Let $P_0$ be
the hydrostatic pressure within the junction. As long as $P_{L,2}>P_0$
the meniscus in branch 2 is arrested. In the following we will answer
the question how long the meniscus in branch 2 will be arrested and we
will implicitly assume that it does not get annihilated by the
filling of the pore from its other end. This means that we assume the
radius $r_2$ also to be larger than the radius of the other branch of
the junction of the other end. This reduces only the probability of
this event by a $r_2$-dependent factor. 

$P_0=P_0(t)$ is a function of time and depends on how far the front
has propagated and can be estimated as follows: Let the average front
height be $\langle h(t)\rangle$. On average one expects the bulk
pressure to decrease linearly from bottom to top:
\be
P(\av{h(t)})/P_0=\av{h(t)}/h_0
\ee
Therefore, with $P(\av{h(t)})=\langle P_L\rangle = -2\sigma\langle
1/r\rangle$ the average Laplace pressure, one obtains
$P_0=-2\sigma\av{1/r}\cdot h_0/\av{h(t)}$ and the condition $P_0=P_{L,2}$
for the arrested meniscus to resume propagation (at time $t_{\rm resume}$) 
reads
%
\be
\av{h(t_{\rm resume})}=h_0 r_2 \av{1/r}\;.
\label{hresume}
\ee
This equation has far reaching consequences:\\ 
1) The larger $r_2$ 
the longer the meniscus is arrested, and the average height that the
front has to reach before the meniscus resumes propagation is
proportional to the height where it stopped with a proportionality
constant larger than one.
2) The time $\tau$ for which the meniscus is arrested is
proportional to the time $t_{\rm stop}$, when it stopped
\be
\tau\propto t_{\rm stop}\;.
\label{tau}
\ee
To see this we note that with (\ref{hresume}) one has
$\av{h(t_{\rm stop} +\tau)}=h(t_{\rm stop})r_2\av{1/r}$.
With Lucas-Washburn $\av{h(t_{\rm stop}+\tau)}
\propto (t_{\rm stop}+\tau)^{1/2}$ and assuming that 
$h(t_{\rm stop})\propto t_{\rm stop}^{1/2}$, too, for the relation
between the height and the time when the considered meniscus stopped,
one obtains (\ref{tau}).\\
3) Consequently from (\ref{tau}) 
\be
\tau \propto h^2(t_{\rm stop})=h_0^2;,
\label{tauH2}
\ee
which implies that the probability distribution of arrest times
for menisci arrested at height $h$ will scale as 
\be 
p_h(\tau)=h^{-2}\,\tilde{p}(\tau/h^2)\;.
\label{ptau}
\ee
4) The height difference $w_0(t_{\rm resume})=\av{h(t_{\rm
resume})}-h_0$ is a measure for the local width of the propagation
front (at the lateral coordinates of the position of the arrested
meniscus) at time $t_{\rm resume}$. The ratio of this local width and
the average height is 
$\frac{w_0(t_{\rm resume})}{\av{h(t_{\rm resume})}}
=1-(r_2\av{1/r})^{-1}$,
which is independent of the time $t_{\rm resume}$.
Thus all arrested menisci will contribute a time independent amount 
to the ratio of the average width $w(t)$ and average height. Since 
the width cannot grow faster than $h(t)$ this implies
\be
w(t)/\av{h(t)}={\rm const.}\;,
\label{wtoh}
\ee
implying $w(t)\propto t^{1/2}$, i.e.\ a roughening exponent
$\beta=1/2$. The constant in (\ref{wtoh}) depends on the pore radius
distribution via the ratio of the minimal and maximal pore radius and
approaches one for an unbounded radius distribution (i.e. the front
extends over the whole occupied volume).

Note that the invasion front dynamics is now expected to be be
completely determined by the meniscus arrests, which in turn depend
exclusively on the pore radii distribution and  the height dependent
hydrostatic pressure. Consequently one expects no lateral correlations
in the meniscus heights to emerge, as observed in \cite{Gruener12}.

The scaling theory presented here neglects all geometric and
topological details of a pore network. To test its
predictions, in particular the strongest (\ref{ptau}) and
(\ref{wtoh}), we analyzed the following microscopic model for
spontaneous imbibition in a pore network with elongated pores
\cite{Lam00,Aker98,Gruener12}: A two-dimensional square lattice of
cylindrical capillaries inclined at $45^{\circ}$ is considered, which
consists of $N_x$ and $N_y$ nodes in horizontal and vertical
directions, respectively. Capillaries, interconnected at nodes, have
the same length $L$ and random radii uniformly distributed
over $[r_{\rm av}-\delta,r_{\rm av}+\delta]$. The average aspect
ratio $2r_{\rm av}/L$ is set to 5.
The pressure at the bottom nodes attached to the liquid
reservoir is set to zero, the pressure at a moving meniscus is
the Laplace pressure.
Here we neglect gravity, which is justified
as long as capillary forces are much larger than gravitational 
forces $2\sigma/r \gg \rho N_y L$, where 
$\rho$ the specific weight of the liquid. This is 
the case for instance in experiments with NVG \cite{Caupin08}.

\begin{figure}[t] \center
\includegraphics[width=1\linewidth]{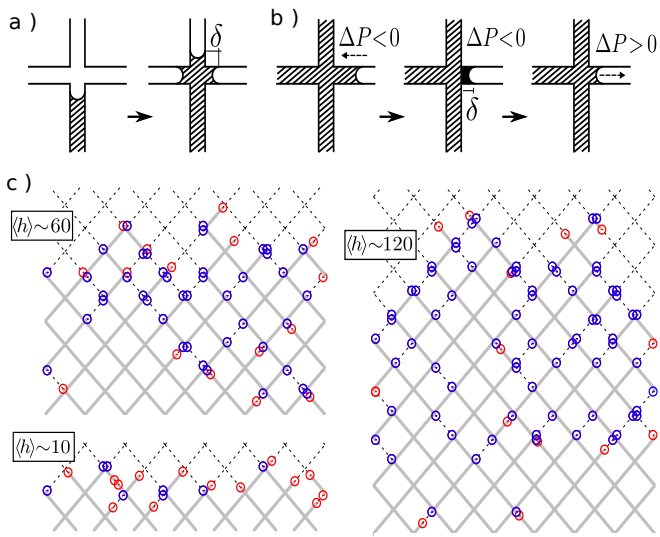}
\caption{{\bf (a,b)}
The mechanisms of menisci advancement in the pore-network model
(a) after reaching a node, and (b) during backward motion and arrest 
of a meniscus due to negative pressure difference. 
{\bf (c)}
Snapshots of the arrested
(blue circles) and advancing (red circles) menisci in the invasion
front at three different times. Broken (full) lines represent empty
(full) pores, $H$ is the average height at the corresponding time.
} 
\label{fig1}
\label{snapshots}
\end{figure}

The hydrostatic pressures at the nodes of the network drives the
dynamical evolution of the menisci configurations. To calculate the
temporal change of the filling heights in the partially filled
capillaries, one needs to know the node pressures which themselves
depend on the menisci configuration. The node pressures $P_i$ are
determined by the boundary conditions plus the conservation of volume
flux at each node: $\sum_j Q_{i}^{j}{=}0$, which is equivalent to
Kirchhoff's law. Here, $Q_i^j$ is the volume flux flowing from node
$i$ into the capillary $j$ attached to it. The sum runs over all of
the four capillaries of node $i$ and is valid for all wet nodes in the
system. According to Hagen-Poiseuille's Law
\cite{ViscousFlow}, $Q_i^j{=}-c_i^j\,\Delta P_i^j / h_i^j$, with
$c_i^j{=}\pi(r_i^j)^4/8\eta$ and $\Delta P_i^j{=}P_i{-}P_{L,i}^j$.
Here, $r_i^j$, $h_i^j$ and $P_{L,i}^j$ are the radius, the length and
the Laplace pressure of the meniscus in capillary $j$ of node $i$,
respectively, and $\eta$ is the viscosity of the liquid. By
numerically solving the resulting set of linear equations we compute
$P_i$ and thus $Q_i^j$. These are then inserted into the equation of
motion for the heights given by $Q_{i}^{j}{=}\pi
(r_i^j)^2\,dh_i^j/dt$. To integrate these differential equations an
implicit Euler scheme with variable time step $\Delta t$ is employed
giving the new positions $h_i^j$. 
When a meniscus reaches the end of a capillary it immediately moves an
infinitesimal distance $\delta{\simeq} 0.01 L$ into the adjacent
capillaries, creating new menisci, as shown in
Fig.~\ref{fig1}(a). This avoids the microscopic treatment of the
filling of the junction \cite{Shikhmurzaev12} and is valid as long as
the filling time of the node is negligible. The filling time
was estimated in \cite{Shikhmurzaev12} and is indeed orders of
magnitudes smaller than the meniscus arrest times 
as long as capillary forces are much larger than
gravitational forces.
When two menisci meet, they vanish, thus the capillary is entirely
filled. If, due to a negative pressure difference, a meniscus
retracts, it proceeds backward as long as its distance from the back
node is larger than $\delta$ [see Fig.~\ref{fig1}(b)]. When it reaches
$\delta$, the meniscus is stuck there until the driving pressure
difference is again positive. During this arrest time, the pressure
calculation is modified with the corresponding capillary being
blocked. We made sure that the simulation results we present in
the following are independent of the choice of $\delta$.

Figure \ref{snapshots} shows three snapshots of the propagating and
arrested menisci in the invasion front at three different times. The
fraction of arrested menisci grows fast with increasing height and
approaches one around $\langle h\rangle\approx 500$.  

\begin{figure}[t] \center
\includegraphics[width=1\linewidth]{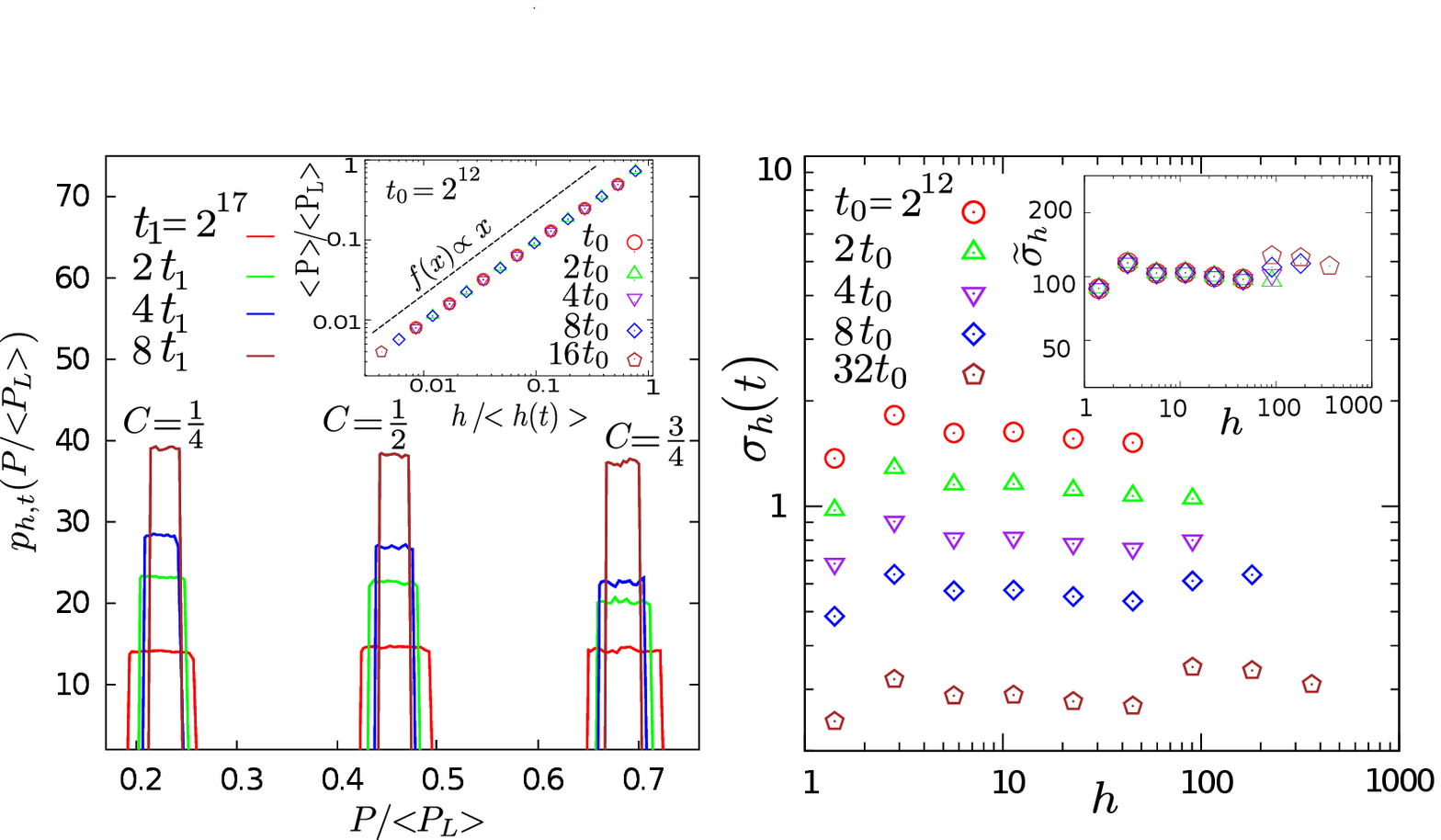}
\caption{
{\bf (a)} Probability distribution $p(P)=p_{h,t}(P)$ of the pressure at
junctions at height $h$ and time $t$. Heights are chosen such that 
for a given time $C:=h/\langle h(t)\rangle$ is constant, data are
shown for $C=1/4$, $1/2$, and $3/4$ and different times.
Inset: Average pressure in junctions at height $h$ at different times.
{\bf (b)} Variance of the pressure distribution for different times as a function of $h$. 
Inset: Scaling plot, $\tilde\sigma_h=\sigma_h(t)/\langle h(t)\rangle$ 
vs. $h$. For all data  $N_x=16$.
}
\label{pressure}
\end{figure}
First we checked the essential assumptions underlying our mean field
description of the imbibition process, namely that the pressure in a
junction can be approximated by the average of the pressure field
$P{\approx}\langle P_L\rangle{\cdot} h/\langle h(t)\rangle$.
Figure \ref{pressure} (a) shows the probability distribution $p_{h,t}(\langle P\rangle/\langle P_L\rangle)$ of the
pressure in the junctions at height $h$ and time $t$.  For a fixed
time we have chosen the height such that the ratio $h/\langle
h(t)\rangle{=:}C$ is constant ($C{=}1/4$, $1/2$, $3/4$ corresponding to
the bottom, middle and upper third of the system). One sees that the
distribution of $P/\langle P_L\rangle$ is centered around $C$,
reflecting that the average pressure indeed is given by
$\langle P\rangle{=}C{\cdot}\langle P_L\rangle$ (see inset), and that the width
systematically shrinks with $t$. The width, given by the 
variance of the pressure distribution, 
$\sigma_h(t){=}(\langle P^2\rangle_{h,t}{-}\langle P\rangle_{h,t}^2)$,
is analyzed in Fig.~\ref{pressure} (b). The inset shows that it
scales as
\be
\sigma_h(t)=\widetilde\sigma_h / \langle h(t)\rangle
\stackrel{t\to\infty}{\longrightarrow} 0\;.
\ee
Consequently the pressure distribution becomes increasingly
sharp with increasing time, which implies that neglecting
pressure fluctuations is a good approximation and leads
asymptotically to correct results.
\begin{figure}[t] \center
\includegraphics[width=1\linewidth]{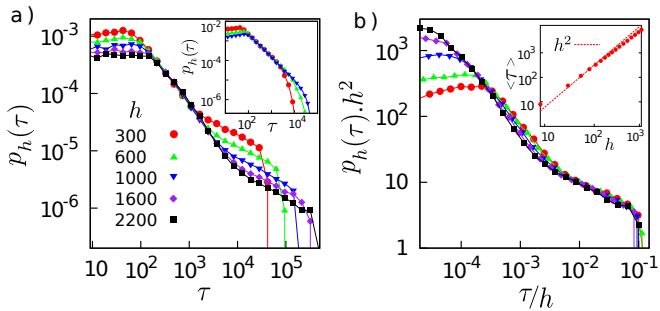}
\caption{{\bf (a)} The probability distribution $p_h(\tau)$ of the 
arrest times $\tau$ of menisci arrested at fixed height $h$ in log-log
scale. Inset: arrest time distribution  of  single meniscus in a pore.
 {\bf (b)} Scaling plot according to eq.(\ref{ptau}). Inset:
Average arrest time $\langle\tau\rangle$ at a given height $h$, dashed
line proportional to $h^2$, see eq.(\ref{tauH2}). Inset: Probability
for a meniscus to be arrested at height $h$. Parameter for all data
are $\delta_r/r=0.1$, $N_x=8$, 100 disorder realization.}
\label{TauB}
\end{figure}

By counting the number of menisci arrested at height $h$ for a time
$\tau$ we obtained the arrest time distribution $p_h(\tau)$, which is
shown in Fig.~\ref{TauB} (a). Three regimes can be identified: A short
time regime, where $p_h(\tau)$ is nearly constant, an intermediate
time regime extending over ca. 1.5 decades for all heights $h$ with a
slope close to $-1$ in the log-log plot, and a large time regime,
extending over ca. 1 decade in $\tau$ with a slope close to
$-1/2$. Finally the distribution is cut-off at a time proportional to
$h^2$. A closer look at the arrest events shows that the main
contribution for the intermediate regime comes from pores with a {\it
single} (arrested) meniscus in a pore (see inset of Fig.~\ref{TauB}
(a)).  Pores that have (arrested) menisci at both ends cause the large
time regime of $p_h(\tau)$, which decays much slower but has a smaller
amplitude. These pores have statistically a larger radius than pores
with only one arrested meniscus - consequently their probability is
lower but arrest times are longer.

The intermediate and large time regime of the distribution,
including the cut-off, scale nicely with $h^2$ as predicted by
Eq.(\ref{ptau}), 
as is shown in Fig.~\ref{TauB} (b). The intermediate and large arrest
times dominate the mean, which is therefore proportional to $h^2$, as
expected from eq.\ (\ref{tauH2}), see inset of Fig.~\ref{TauB} (b).
Events with a brief arrest time, which make for the short time regime
that does not scale with $h^2$ (see Fig.~\ref{TauB} (b)), are caused
by secondary arrests of menisci and by small radii differences in
adjacent pores and dynamic fluctuations in the node pressures due to
propagation and release of nearby menisci.

We also computed the average height $\langle h(t)\rangle$ and width
$w(t){=}(\langle h^2(t)\rangle-\langle h(t)\rangle ^2)^{1/2}$ of the
imbibition front and found that the ratio $w(t)/\langle h(t)\rangle$
approaches a constant value for large times, which confirms the
prediction (\ref{wtoh}). Since $\langle h(t)\rangle{\propto} t^{1/2}$
the width also increases as $w(t){\propto} t^{1/2}$, implying a
roughness exponent $\beta{=}1/2$. The initial decrease of
$w(t)/\langle h(t)\rangle$ indicates a pre-asymptotic increase of
$w(t)$ with an exponent slightly smaller than $1/2$, as reported in
\cite{Gruener12}. 

The invasion front thus involves a finite fraction of the occupied
volume and comprises connected clusters of empty pores whose size
distribution gets broader with increasing time. Based on the
conditions for meniscus arrests presented above one can derive a
scaling form for the distribution of cluster sizes as follows.
Consider an empty cluster that contains $S$ pores at time $t$ (i.e.\
with width $w(t)$). Its lateral size scales as ${\cal L}\sim
S^{1/d_f}$ and its surface area as ${\cal F}\sim S^{d_s/d_f}$, where
$d_f$ and $d_S$ are the bulk and surface fractal dimension of the
empty clusters ($d_f=d$ and $d_s=d-1$ in the case of compact
clusters). Almost all pores in the boundary ${\cal F}$ of the cluster
have arrested menisci, and for a meniscus to be arrested the radius of
its pore has to be larger than the radius of an adjacent pore, which
is an event that occurs with some probability $q<1$. Assuming that the
conditions for meniscus arrest in all boundary pores are independent
from one another and the boundary consists of the order of
$S^{d_s/d_f}$ pores, the probability for collective meniscus arrests
in boundary pores is proportional to $\exp(-\alpha\cdot S^{d_s/d_f})$,
where the constant $\alpha$ involves $\ln q$ and a geometric
factor. Since the lateral dimension $S^{1/d_f}$ of the cluster must
not exceed the width $w(t)$, one obtains for the probability of an
arbitrary empty pore in the front region to belong to a connected
cluster with $S$ empty pores:
\be
q_S={\cal N}^{-1}\,S\exp(-\alpha\cdot S^{(d_s/d_f})
\cdot\tilde{g}(S^{1/d_f}/w(t))\;,
\label{cluster}
\ee
where ${\cal N}$ is a normalization factor and $\tilde{g}(x)$ is a
scaling function that is $1$ for $x\ll1$ and $0$ for $x\to1$. A
cluster analysis of our simulation of the 2$d$ pore network model confirms
the stretched exponential behavior of $q_S$ at large times ($w(t)\gg
S^{1/d_f}$) with $d_s/d_f$ close to $0.5$ $=(d-1)/d$.

In conclusion we have presented a scaling theory for
the imbibition of an arbitrary wetting liquid through any porous medium 
consisting of random networks of elongated pores. We tested the
predictions in extensive simulations of a pore network model. Meniscus
arrest times at pore junctions are shown to scale with the age of the
invasion front whose width is therefore proportional to its average
height. This establishes a universality class for invasion front
broadening that is realized in nano-porous Vycor glass
\cite{Gruener12} and is expected to determine roughening dynamics in
similar porous media. Since meniscus arrest is solely determined by
the relation of radii of the pores emanating from one junction, it
should be possible to relate dynamical quantities accessible via light
or Neutron scattering to characteristics of the pore radius
distribution of the porous medium.

{\bf Acknowledgements:} We thank P. Huber and D.-S. Lee for 
stimulating discussions.

\end{document}